\documentclass[]{fmj2010}

\pdfoutput=1

\usepackage[pdftex]{graphicx}

\newcommand{\Rs}{R_{\rm S}}
\newcommand{\bs}{\beta_{\rm s}}
\newcommand{\rmerge}{r_{\rm mrg}}
\newcommand{\rcoll}{r_{\rm col}}
\newcommand{\Eint}{E_{\rm int}}
\newcommand{\xint}{\xi_e}
\newcommand{\xiB}{\xi_B}
\newcommand{\gl}{\check\gamma}
\newcommand{\gu}{\hat\gamma}
\newcommand{\Msol}{M_\odot}
\newcommand{\tsyn}{t_{\rm syn}}
\newcommand{\tacc}{t_{\rm acc}}
\newcommand{\Lsyn}{L_{\rm syn}}
\newcommand{\bsh}{\beta_{\rm sh}}
\newcommand{\nua}{\nu_{\rm a}}
\newcommand{\nub}{\nu_{\rm b}}
\newcommand{\nuc}{\nu_{\rm c}}
\newcommand{\num}{\nu_{\rm m}}
\newcommand{\Sm}{S_{\rm m}}
\newcommand{\siT}{\sigma_{_{\rm T}}}

\newcommand{\lsim}{\mathrel{\raise.25ex\hbox{$<$}\kern-.8em\lower.9ex\hbox{$\sim$}}}
\newcommand{\gsim}{\mathrel{\raise.25ex\hbox{$>$}\kern-.8em\lower.9ex\hbox{$\sim$}}}

\title{Radio-submm flares from blazars in a discontinuous jet model}
\author{J\"org P. Rachen\inst{1} \and Max H\"aberlein\inst{1}\thanks{Walther-Mei{\ss}ner-Institut, 85748 Garching} 
	\and Felix Reimold\inst{1}\thanks{Max-Planck-Institut f\"ur Plasmaphysik, 85748 Garching}
	\and Thomas P. Krichbaum\inst{2}
          }

   \institute{Max-Planck-Institut f\"ur Astrophysik, Karl-Schwarzschild-Str. 1, 85748 Garching, Germany
         \and Max-Planck-Institut f\"ur Radioastronomie, Auf dem H\"ugel 69, 53121 Bonn, Germany}

\abstract{We present a model in which AGN jets are described by a dense series of plasma blobs, distributed in masses and velocities. The blobs expand and collide, and ultimately form a continuous jet flow as observed on larger scales. When blobs collide, energetic electrons are produced by first order Fermi acceleration, and emit synchrotron radiation at the radio-optical frequencies with a double-broken power law spectrum. The difference of this model to the well-known Marscher\&Gear-model is, that in our model shocks are produced on a range of scales, and exist only temporarily, which causes differences in the prediction for the spectral evolution of flares. We apply our model to radio-submm data obtained for 3C454.3, and briefly discuss implications for gamma-ray production.}

\begin{document}

\maketitle

\section{Introduction}

Based on the work of \cite{MG85}, flares in AGN jets usually modelled by shocks propagating in a continuous plasma jet from its origin to the outside. While the shock continuously energizes electrons, synchrotron radiation emitted by it goes through a flare, owing to the competing cooling processes of inverse Compton radiation, dominating the in the rise phase, and adiabatic losses in the expanding jet, dominating in the decay phase, with the possibility of a plateau phase in-between, in which synchrotron radiation itself is the dominating cooling process. As the relative strength of cooling processes depends on global jet parameters, which are unlikely to change on a short timescale, all flares should evolve in a similar way, with their rise, plateau and decay timescales being of the same order of magnitude, typically weeks to months. While this fits observations of radio flares quite well, the situation is less clear in the optical and gamma-ray regime. Here, flares appear on a large range of time scales, from less than an hour in some TeV blazars, to weeks for strong blazars in the Fermi regime. 

To accommodate the situation in the gamma-ray regime, \cite{Spada01} suggested a so-called ``internal shock model'' for blazars, motivated by the model of \cite{MR94} for the explanation of rapidly variable gamma-ray burst lightcurves. In this model, shocks are formed in random collisions of shells, and subsequently accelerate electrons to emit synchrotron and inverse Compton radiation. Within the framework of this synchrotron-self Compton (SSC) scenario, \cite{Spada01} used their model to explain qualitatively the global characteristics of blazar variability at optical to gamma-ray frequencies. 

In the current work we revoke the internal shock model for blazars, focussing on the explanation of flare evolution in the radio-submm regime, which we claim to allow us to pinpoint the physical parameters of the jet independent on the assumptions for gamma-ray production. A detailed discussion will be given by \cite{HR10}, henceforth referred to as ``Paper I''. Here we discuss briefly the basic idea and properties of this model, and apply it to a flare observed in the bright blazar 3C454.3 in 2005.

\section{Jets, shocks and plasmons : not a contradiction}

\begin{figure*}
\centering
\includegraphics[width=0.9\textwidth,height=5.2cm,viewport= 0 100 850 480]{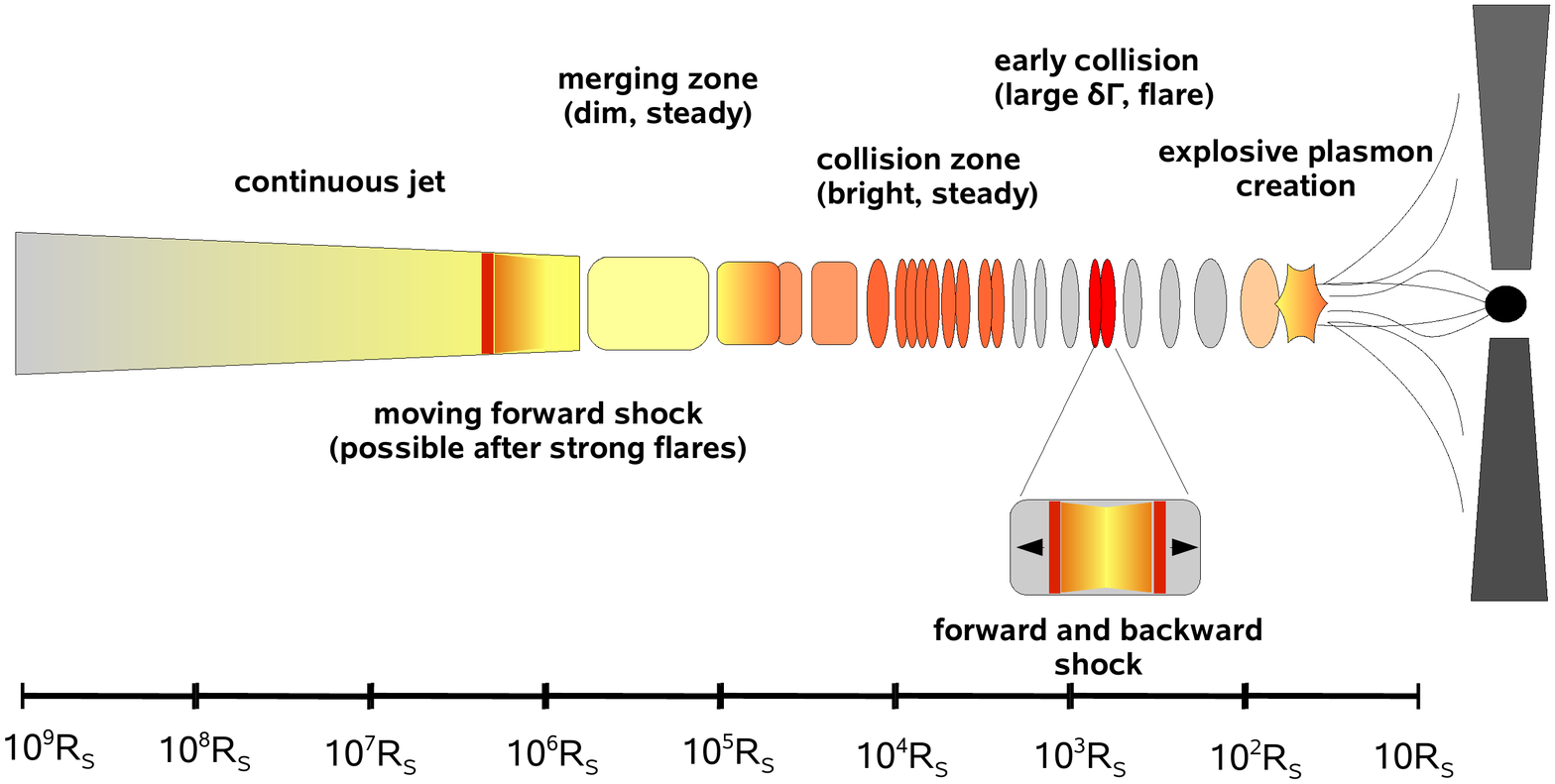}
   \caption{\label{fig:model} \small Illustration of the discontinuous jet model. Plasmons created some $10{-}100$ Schwarzschild-radii from the black hole collide mostly at around $10^4\Rs$, converting a significant fraction of the bulk Lorentz factor variance into internal energy, and ultimately into radiation. As the ``collision region'' is quasi-stationary in position and emits most of the radiation in a spectrum averaging out the individual collisions, it might be identified with the bright ``millimetre-core'' seen in many blazars. Particularly large differences in bulk Lorentz factor may lead to early, strong collisions which appear as flares in the spectrum. As the collision region is leveling out the differences in bulk speed, the jet merges afterwards quite smoothly. A series of fast blobs, e.g. after a surge in the accretion rate, may propagate through the whole system, leading to a series of flares, and ultimately form a moving shock in the continuous jet, explaining VLBI observations as reported by \cite{Marscher08}.}
\end{figure*}

Essentially two fundamental mechanisms have been proposed to explain why plasma jets may be launched from accreting black holes: the \cite{BZ76} process, which is based on electromagnetic extraction of the angular momentum of a rotating black hole in form of a strong Poynting flux, and the \cite{BP82} process of hydromagnetic extraction of rotating plasma from the corona of the accretion disk. In the recent discussion of AGN and other accreting sources, it is generally assumed that both mechanisms contribute to the jet formation, possibly with different weights depending on the physical properties of the accreting system [e.g., \cite{Meier01})]. Although plasma physicists have shown that solutions for the continuous emission of jets exist, all involved physical processes are prone to instabilities, in particular in the highly relativistic regime of accreting black holes. This justifies the assumption that, although jets are observed as continuous plasma flows on larger scales, they are not born like this, but as a dense, random flow of plasma blobs, emitted a central source similarly to a discharge process.

Let us assume that the blobs are emitted on an average rate $f_* < c / \Rs$, if $\Rs$ is the Schwarzschild radius of the central black hole, at highly relativistic speeds with a bulk Lorentz factor $\Gamma_* \sim 10$. Being generally unconfined, in their comoving frame the blobs would expand in all directions approximately with their speed of sound, $\bs c$, leading to an opening angle of the yet discontinuous jet of $\bs\Gamma_*^{-1}$. In the initial phase, where relativistic gas dominates the pressure, the expansion decreases with distance $r$ from the black hole approximately as $\bs \propto r^{-1/2}$, leading to an apparent collimation of the jet to angles of less than $1^\circ$. After this, we assume conical expansion with $\bs\sim 0.1$. Neglecting velocity differences of the blobs, they would merge rather smoothly after a distance $\rmerge \sim \Gamma_*^2 c / f_* \bs$, as their expansion has filled up the gaps between them. Of course, a realistic ``discharge'' process near the black hole will emit the blobs with randomly distributed Lorentz factors $\Gamma_* + \delta\Gamma_i$, with $\langle\delta\Gamma_i\rangle = 0$ and a normalized standard deviation $\sigma_\Gamma = \sqrt{\langle\delta\Gamma_i^2\rangle} / \Gamma_*$. Inelastic collisions between plasmons will then become frequent at a distance $\rcoll \sim \Gamma_*^2  c / f_* \sigma_\Gamma$, converting on average a fraction $\sigma_\Gamma^2/4$ of the bulk flow into internal energy. %
Subsequent collisions will lead to more and more elongated ``jet-bits'', with the spread in Lorentz factor diminishing as ${\sigma_\Gamma}^{[k]} \sim {\sigma_\Gamma}^{[0]}/2^k$, if $k$ denotes the number of collisions an individual blob has suffered. Eventually, ${\sigma_\Gamma}^{[k]}\ll\bs$, and blob-expansion is again the dominating process to form the continuous jet at $\gsim 10^5\Rs$. Figure \ref{fig:model} illustrates this process. 

We want to emphasize that our plasmons cannot be identified with the observable VLBI components. In our model, there are $\sim\Gamma_*^2 / \bs \sim 10^3$ plasmons in the discontinuous phase of the jet ($r \lsim 10^5\Rs$) at any time. They are usually dark, as they are born with extremely high magnetic field cooling down any synchrotron emission on a short timescale. The blobs become bright only after they are re-energized at $\sim 10^4{-}10^5\Rs$ in collisions. For most blazars, even the best VLBI pictures will not be able to resolve single collisions; rather we expect that the known resolved VLBI features in AGN jets correspond to the distinct zones in our model, as shown and explained in Fig.~\ref{fig:model}. We also emphasize that, in general, collisions in our model do not correspond to a ``flares'' in the observed spectrum. As the rate of collisions must be equal to $f_*$ to create a stationary system, that means for $f_*\sim c/10 \Rs$ and a $10^8\Msol$ black hole several collisions per day. As the synchrotron cooling time scale is much larger, i.e. $f_* \tsyn \gg 1$, we see at any instant the superposed emission of many plasmon collisions, averaging out their spectral and temporal properties. The average collisions with $\delta\Gamma/\Gamma \lsim \sigma_\Gamma$ therefore represent what observes call the \textit{quiescent} emission of a blazar; \textit{flares}, in the contrary, are collisions with $\delta\Gamma/\Gamma \gsim 3\sigma_\Gamma$, which can have luminosities up to three orders of magnitude higher than the average, as we will show below. In Paper I we present a full scale simulation of the jet, starting from the statistical properties of the central engine and calculating of all $n$-th order collisions on the way to the continuous jet. There we also we demonstrate that the overlay of average ``mini-flares'' indeed produces a broken power law spectrum with $\alpha\approx 0$ below, and $\alpha\approx 0.7{-}1.2$ above the break frequency typically in the range $10{-}300\,$GHz, as usually observed in blazars.

\section{Flare characteristics in a discontinuous jet}

When two plasma blobs, both carrying an energy $\approx M c^2 \Gamma$, collide with a Lorentz factor difference $\delta\Gamma$, two shocks running in opposite directions with speeds $\bsh\approx\sqrt{2 \delta\Gamma/\Gamma}$ (for $\delta\Gamma\ll\Gamma$) are formed, and internal energy $\Eint \sim \frac14 M c^2 (\delta\Gamma/\Gamma)^2$ is created. We assume that a significant part of this internal energy is going into turbulence, specifically magnetic field creation in turbulent dynamos, and another significant part into nonthermal particles through the mechanism of first order Fermi acceleration at the shocks. First order Fermi acceleration has the property to \textit{produce} power law particle distributions, and therefore to \textit{explain} the ubiquitous presence of power law spectra in sources known to contain shocks (see \cite{Drury83} and references therein). We assume in the following the canonical power law index for Fermi-accelerated electrons, i.e.\ $N(\gamma) = N_e \gamma^{-s}$ with $s=2$, for the electron Lorentz factor in the range $\gl\le\gamma\le\gu$. We introduce the parameters $\xiB$ and $\xint$ by $u_B \equiv B^2 / 8\pi = \xiB \Eint/2 V$ and $u_e = N_e m_e c^2 \ln(\gu/\gl) = \xint\Eint/2 V$, where $V \approx \frac43\pi R^3$ is the volume of a ``virgin'' plasmon. In a conical jet, $R$ related to the distance $r$ from the black hole as $R \approx r \bs / \Gamma$. As the synchrotron luminosity $\Lsyn\propto R^2 N_e B^2$, we get $\Lsyn\propto \Eint^2/R^4 \propto (\delta\Gamma/\Gamma)^8$ for constant $\gu/\gl$.\footnote{\cite{Spada01} vary $\gl$ rather than $N_e$ to account for constant fraction of nonthermal energy in electrons to total internal energy, which would lead to $\Lsyn\propto (\delta\Gamma/\Gamma)^3$.} Therefore, ``$3\sigma$-events'' in the distribution of collisions can outshine the sum of all average collisions, revealing temporarily the spectral properties of a single emission region with a specific size $R$, and specific values for plasma properties like $N_e$ or $B$. This is, what we usually call a \textit{flare}. 

The spectrum of a flare is represented as a broken power law with three characteristic frequencies. The first is the synchrotron-self absorption frequency $\nua$, defined by requiring that the opacity $\tau_{\rm ssa}(\nua)=1$, yielding
\begin{equation}\label{nua}
\nua \approx \left[\frac{\siT R u_B u_e}{4 m_e c^2 \ln(\gu/\gl)}\right]^{\frac13}\quad.
\end{equation}
where $\siT = 8\pi r_e^2/3$ is the Thomson cross section. For $\nu<\nua$, the spectrum is $S_\nu \propto \nu^{5/2}$, for $\nu>\nua$ we will have $S_\nu \propto \nu^{-\alpha}$ with $\alpha=(s-1)/2=0.5$. This spectral index, however, is only correct for $\tsyn(\nu) > t$, i.e., synchrotron cooling is not yet efficient for the electrons emitting at a characteristic frequency $\nu(\gamma) = 3 e B \gamma^2/ 16 m_e c$ at a time $t$ after the collision. Above a $\tilde\nu_{\rm b}(t)$ for which $\tsyn=t$, the electron spectrum steepens as $s\to s+1$, yielding $S_\nu \propto \nu^{-1}$. Evaluating $\tilde\nu_{\rm b}(t)$ for the fully developed flare, i.e., $t \approx R / c \bsh$, we obtain
\begin{equation}\label{nub}
\nub \approx \frac{R c \bsh^2}{r_e^2}\left[\frac{m_e c^2}{e B R}\right]^3
\end{equation}
Finally, the spectrum reaches a maximum frequency at $\nuc = \nu(\gu)$, above which it turns into an exponential cutoff. The value of $\gu$ is hereby determined by the relation $\tacc(\gu) = \tsyn(\gu)$, where $\tacc(\gamma) \approx 3\kappa/c^2 \bsh^2$ is the acceleration time scale for first order Fermi acceleration. \cite{BS87} connected the diffusion coefficient $\kappa$ to the plasma turbulence spectrum, and showed that for a Kolmogoroff spectrum known from fully developed hydrodynamical turbulence, $\nuc$ is expected in the infrared to optical regime, as observed for many AGN. In Paper I, we will discuss a slightly modified version of the BS87 approach in more detail; here, we focus on the break frequencies $\nua$ and $\nub$, which are more relevant in the radio-submm regime.

We evaluate $\nua$ and $\nub$ numerically for the ``average'' collision, i.e., with $R=\rcoll\bs/\Gamma_*=\Gamma_* c\bs/f_*\sigma_\Gamma$, $\bsh = \sqrt{2\sigma_\Gamma}$, for  values $\bs=0.1$, $f_*=c/100\Rs$, and $\ln(\gu/\gl)=7.5$. If we further assume that the jet power is equal to the Eddington luminosity of the black hole, with mass $10^9\,M_9\,\Msol$, we find
\begin{eqnarray}
\nua' &\sim& 8\;10^{13}\,{\rm Hz}\;\sigma_\Gamma^3\,\xiB^{1/3}\,\xint^{1/3}M_9^{-1/3}\!\!\!\!\rightarrow 200\,{\rm GHz}/200\,{\rm GHz}\\
\nub' &\sim& 1.5\;10^6\,{\rm Hz}\;\sigma_\Gamma^{-9/2}\xiB^{-3/2}M_9^{-1/2} \rightarrow 150\,{\rm GHz}/1{\rm GHz}
\end{eqnarray}
We defined $\nu'=D\nu$ to consider that blazar emission is usually Doppler boosted, setting $D=\Gamma_*=10$. The values after the arrow correspond to $\sigma_\Gamma=0.3$ and $M_9=5$ (as expected for 3C454.3), the first value for $\xiB=0.3$ and $\xint=0.01$, and the second for $\xiB=0.01$ and $\xint=0.3$ (see Section~\ref{sec:gamma} for a discussion). We see that for all cases $\nub<\nua$, that means we expect in flares a steep spectrum $\propto \nu^{-1}$ above $\nua$. Note, however, that during the rise phase, $\tilde\nu(t)>\nub$, so that the flare may temporarily exhibit a $\nu^{-0.5}$ spectrum directly above $\nua$. 

Of particular interest is the behavior of the flux at $\nua$ during the flare evolution, as this corresponds to the $(\num,\Sm)$ plots often used in radio astronomy to describe flare evolution. In our model, $B$ is approximately constant while the shock is active, and $R$ hardly changes during the shock crossing time. Therefore, we expect that $\Sm$ rises for constant $\num$.  When the flare decays adiabatically after the shock ceases, the flux in the optically thick region evolves as $S_\nu \propto R^2 B^{-1/2}$, as the source function does not depend on the electron spectrum. With $S_\nu\propto \nu^{5/2}$ and for $B\propto R^{-2}$, one obtains $\Sm\propto \num^{1.2}$ (one can easily check that $\Sm\propto\num^1$ is obtained for $B\propto R^{-1}$). If synchrotron cooling dominates over adiabatic cooling at $\nua$, which is normally the case initially as $\nub\ll\nua$, one can show that $\Sm\propto\num^{5/2}$, i.e., the flare decays along the spectral edge of self-absorption. This may be compared with the predictions of the \cite{MG85} model, which are quite different. A more detailed discussion of flare evolution in our model will be given in Paper I. 


\section{Fitting the 2005 flare in 3C454.3}
\label{sec:fit}

To illustrate the capabilities of our approach to reproduce the temporal-spectral evolution of blazar flares, we apply our model to data obtained on the bright blazar 3C454.3 in a campaign which has been started in 2005 after an initial strong mm-/optical flare. The main goal of the campaign has been to obtain quasi-simultaneous radio-spectra covering the frequency range from 1 - 300\,GHz, which can be used to study the spectral evolution over time. 


The Effelsberg 100\,m radio-telescope of the Max-Planck-Institut f\"ur Radioastronomie with its sequence of secondary focus heterordyne receivers is well suited to measure flux densities quasi-simultaneously and within 30-40 mins at $2.7$, $5$, $8.4$, $10.5$, $15$, $22$, $32$, and $43$\,GHz. These measurements were occasionally complemented by flux density measurements at $0.6$-$1.4$\,GHz, when the corresponding prime-focus receivers were available. The data were analyzed and calibrated in the standard way following similar procedures as described in \cite{Fuhrmann08} and \cite{Angelakis08}. The Effelsberg radio data were complemented by measurements within a few days of the millimeter flux densities obtained through the regular AGN monitoring program, which is performed at the IRAM 30\,m telescope on Pico Veleta (Spain). This instrument provided measurements at $90$\,GHz and $230$\,GHz (see \cite{Ungerechts98} and \cite{Agudo06} for more details and a description of the data reduction). We further included $230$\,GHz and $345$\,GHz data from the AGN-monitoring of the sub-millimeter array (SMA) on Mauna Kea (Hawaii, see \cite{Gurwell07} and refs.\ therein), which are published in part by \cite{Villata09} and refs.\ therein.


\begin{figure}
\centering
\includegraphics[width=\columnwidth, height=6cm, viewport=50 50 760 550]{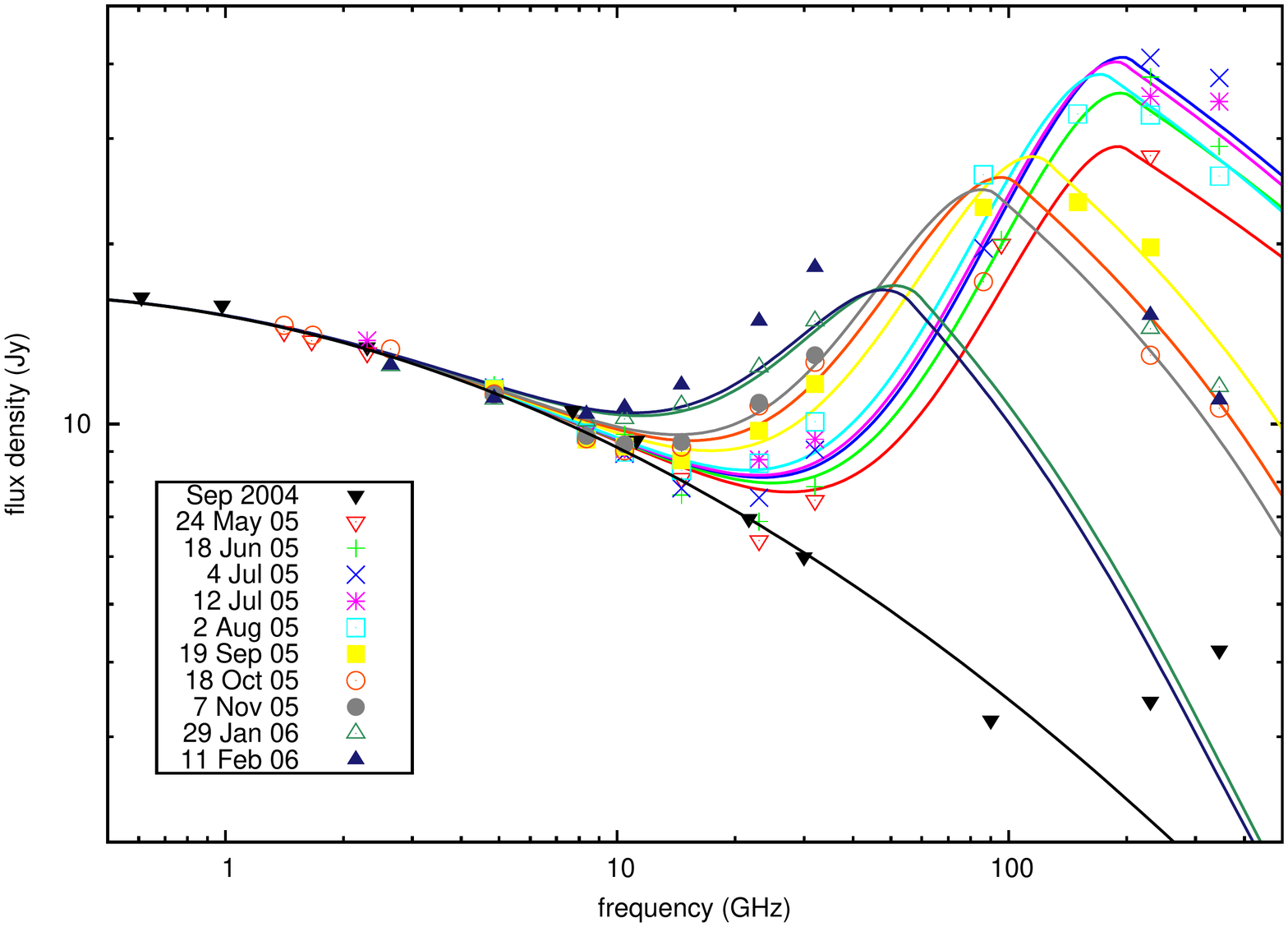}
\caption{\small Fit of a single flare in the discontinuous jet model to data of the 2005/06 flare from 3C454.3, see also \cite{Krichbaum06} and the discussion in the text. \label{fig:3C454}
}
\end{figure}

Figure \ref{fig:3C454} shows the data together with spectra simulated from our full model, as described in Paper I. We assumed that two plasmons with $M c^2 = 7\;10^{54}\,$erg, $\Gamma_1=10$ and $\Gamma_2=13$ collide at a distance ${\sim}10^{19}\,$cm from the black hole, consistent with a jet power of $10^{48}\,{\rm erg/sec}$ and $f_* \sim 10^{-6}\,$Hz. We further assumed $\bs=0.1$, thus $R\sim 10^{17}\,$cm, and a constant viewing angle of $5^\circ$ to calculate the Doppler factor. All parameters have then been optimized in an adaptive method by minimisation of $\sum_t\sum_i\chi^2_{i,t}$ for all data sets taken at the times $t$ as given in the plot, where the simulated flare spectra have been added to a constant underground fitted to the data from September 2004. As the starting time of the flare is unknown, we have identified the data set from July 4th with the maximum of the flare. Frequencies and times have been properly corrected for the redshift of the source, $z=0.859$. As the best fit parameters, we found $\xiB\approx0.3$, and $\xint\approx 10^{-3}$. The temporal evolution of the flare is represented quite well, until presumably a new flare arises beginning of February 2006. The discrepancies at the low frequency turnover from the background to the flare spectrum should not surprise, as our model expects the background to change with time. Clearly, data at frequencies above $350\,$GHz would have been helpful to further constrain our parameters.

\section{Implications for gamma-ray emission}
\label{sec:gamma}

In their discontinuous jet model, \cite{Spada01} included inverse Compton (both SSC and external IC) emission, and determine the necessary jet parameters from the requirement to fit existing gamma-ray data. In contrast, it is our approach to determine the jet parameters from synchrotron emission alone, and keep minds open regarding the mechanism of gamma-ray production.  We may remind the readership at this point that there are alternatives to inverse Compton models for gamma-ray production. In the so-called ``hadronic'' scenario, protons are accelerated to extremely high energies, enabling them to produce gamma-rays by pion induced cascades [\cite{Mannheim93}], with a possible admixture of proton-synchrotron radiation [\cite{Muecke03}]. As first order Fermi acceleration, which is the base of our model, does not distinguish between protons and electrons in principle, we may state that hadronic models are at least as ``natural'' as the ``leptonic'' inverse-Compton scenario. However, hadronic models are very difficult to handle technically, so we abstained from making detailed predictions for the gamma-ray spectra in our model at first; nevertheless it is possible to draw some preliminary conclusions on gamma-ray production from our results.

As discussed by \cite{Rachen00}, leptonic and hadronic models can be distinguished by the required relation of magnetic field to photon density, i.e., by the ratio $\xiB/\xint$: while leptonic models need $\xiB\lsim \xint$ (in case of TeV blazars even $\xiB\ll \xint$) to explain gamma-ray data, hadronic models require $\xiB \sim \xi_p \gg \xint$, where $\xi_p$ denotes the fraction of internal energy in relativistic protons. For our fit of the flare of 3C454.3, $\xiB\gg \xint$ was found. It would be premature to consider this evidence for hadronic origin of gamma rays from 3C454.3, as our fit method does not really perform a full parameter space search, identifying all allowed and excluded regions. However, it does give confidence in our method to distinguish between hadronic and leptonic models in future, improved applications. This certainly requires to include \textit{all} gamma-ray production methods into our model, and then compare with \textit{both} Fermi gamma-ray, and \textit{detailed} radio-optical data. For the latter, as obvious from the discussion in Section~\ref{sec:fit}, closing the infrared gap in regular blazar monitoring would be highly desirable.

\end{document}